\documentstyle[12pt,epsfig]{article}
\def\lapp{{\ \lower 0.6ex \hbox{$\buildrel<\over\sim$}\ }}
\def\gapp{{\ \lower 0.6ex \hbox{$\buildrel>\over\sim$}\ }}
\begin{document}
\begin{titlepage}
\vspace*{-1cm}
\begin{flushright}
DTP/96/04  \\
January 1996 \\
\end{flushright}                                
\vskip 1.cm
\begin{center}                                                                  
{\Large\bf Non--minimal neutral Higgs bosons at LEP2 }
 \vskip 1.cm
{\large A.G.~Akeroyd\footnote{A.G.Akeroyd@durham.ac.uk}}
\vskip .4cm
{\it Department of Mathematical Sciences, Centre for Particle
Theory,\\ University of Durham, \\
Durham DH1 3LE, England.}\\
\vskip 1cm                                                                    
\end{center}

\begin{abstract}
We study the phenomenology of the neutral Higgs sector of a non--SUSY
non--minimal Standard Model. Models with more than one Higgs doublet
are possible, and may contain neutral Higgs
scalars with branching ratios significantly different to those of the 
Minimal
Standard Model Higgs boson. We show how these differences may be exploited
at LEP2 in order to distinguish the non--minimal Standard Model from the
minimal version.

\end{abstract}
\vfill
\end{titlepage}                                                                 
\newpage       

\section{Introduction}
It is well known that the Standard Model (SM) \cite{Wein} requires the breaking
of the $SU(2)\times U(1)$ symmetry. Introducing a complex scalar doublet
with a non--zero vacuum expectation value (VEV) is an elegant way of achieving
this, and predicts one neutral scalar -- the Higgs boson ($\phi^0$)
 \cite{Higgs}. Models 
with $N$ doublets, which we shall call `multi--Higgs--doublet models' (MHDM) 
are possible \cite{Gross}, 
and in particular two doublets are required for the minimal Supersymmetric 
Standard Model (MSSM) \cite{Gun}. One of the main goals at LEP2 will be to 
search for  
the above Higgs bosons, with previous searches at LEP having produced bounds
of $M_{\phi^0}\ge 65.1$ GeV \cite{Sop} for the minimal SM, and $M_{h^0}\ge 44$ 
GeV \cite{Del} for the
lightest CP-even neutral Higgs boson ($h^0$) of the MSSM. 

Although there are
many theoretically sound extended Higgs models, most attention is given to
the minimal SM and the MSSM. In this paper we study the various 
non--supersymmetric non--minimal Higgs models in order to ascertain whether 
distinctive signatures 
are possible. We shall focus on the lightest CP--even neutral scalar ($h_1$)
and its 
phenomenology at LEP2, assuming that this Higgs particle is the only one
in range at this collider. If a 
neutral Higgs boson is detected at LEP2 it is important to know from which 
model it originates. We shall only consider models with Higgs isospin doublets 
and our work is organized as follows. In Section 2 we introduce the relevant 
non--minimal Higgs models and investigate their couplings. Section 3 examines 
their phenomenology at LEP2 in the case of distinctive branching ratios not
possible for $\phi^0$ and $h^0$. Finally, Section 4 contains our conclusions.
         
\section{Extended Higgs Sectors}

The minimal SM consists of one Higgs doublet ($T=1/2$, $Y=1$), although 
extended sectors can be considered and have received substantial 
attention in the literature. For a general review see Ref. \cite{Gun}. There 
are two main constraints on such models:
\begin{itemize}
\item[{(i)}] There must be an absence of flavour changing neutral 
currents (FCNC).
\item[{(ii)}] The rho parameter, $\rho=M^2_W/(M^2_Z\cos^2\theta_W)$, must 
be very close to one. 
\end{itemize}
Condition (i) is satisfied by constraining the Yukawa couplings to the 
fermions \cite{Wein2}. Condition (ii) requires models with only doublets, to 
which any number of  singlets ($T=0$, $Y=0$) can be added. Models with 
triplets ($T=1$) \cite{Geo}, \cite{Veg}, \cite{Ake2} can also
be considered, although obtaining $\rho\approx 1$ is achieved in a less 
natural way than for cases with only doublets. 

The theoretical structure of the two--Higgs--doublet model (2HDM) is well 
known \cite{Gun}, while the general multi--Higgs--doublet model (MHDM) 
 \cite{Gross}, \cite{Bran}, \cite{Ake} 
has received substantially less attention. In particular when the MHDM is 
considered in the literature the focus is nearly always on the charged Higgs 
sector. We shall be investigating the properties of the lightest neutral 
CP-even Higgs boson of the MHDM, and for simplicity assume no mixing
between CP-even and CP-odd scalars. It has been shown \cite{Gross} that 
significant differences can exist between the 2HDM and MHDM in the case of
the charged Higgs sector, and we anticipate an analogous result in the 
neutral sector.

\begin{table}[htb]
\centering
\begin{tabular} {|c|c|c|c|c|} \hline
 & Model~I & Model~I$'$ & Model~II & Model~II$'$  \\ \hline
u (up--type quarks)   & 2 & 2 & 2 & 2 \\ \hline
d (down--type quarks) & 2 & 2 & 1 & 1 \\ \hline
e (charged leptons)   & 2 & 1 & 1 & 2 \\ \hline
\end{tabular}
\caption{The four distinct structures of the 2HDM.}
\end{table}

Table 1 shows the four different ways with which the 2HDM can be coupled to 
the fermions (the Yukawa couplings). The numbers (1 or 2) show which Higgs 
doublet couples to 
which fermion type. The Higgs sector of the MSSM requires Model~ II type 
couplings and thus the phenomenology of Model~ II has received the most
attention of the four. Models I$'$ and II$'$ are rarely 
mentioned, and only then in the context of the charged Higgs sector \cite{Phil}.
For a MHDM there are more permutations of the Yukawa couplings, although it is 
conventional to couple each fermion type to a distinct doublet, e.g. in a 
3HDM doublets 1, 2, 3 would couple to d, u, e
respectively. We shall be considering the
lightest CP-even Higgs scalar ($h_1$) of the above models, and for the 2HDM 
its couplings to the fermions are given
in Table 2.\footnote{For Higgs masses in the range of LEP2, branching ratios  
to vector bosons are negligible.}   

\begin{table}[htb]
\centering
\begin{tabular} {|c|c|c|c|c|} \hline
& Model~I & Model~I$'$ & Model~II & Model~II$'$  \\ \hline
$hu\overline u$ & $\cos\alpha/\sin\beta$ & $\cos\alpha/\sin\beta$
& $\cos\alpha/\sin\beta$  & $\cos\alpha/\sin\beta$      \\ \hline
$hd\overline d$ & $\cos\alpha/\sin\beta$& $\cos\alpha/\sin\beta$ 
&$-\sin\alpha/\cos\beta$ &$-\sin\alpha/\cos\beta$      \\ \hline
$he\overline e$ &  $\cos\alpha/\sin\beta$  & $-\sin\alpha/\cos\beta$ 
  &$-\sin\alpha/\cos\beta$ 
&  $\cos\alpha/\sin\beta$\\ \hline
\end{tabular}
\caption{The fermion couplings of $h_1$ in the 2HDM relative to those for the
 minimal SM Higgs boson ($\phi^0$).}
\end{table}
\noindent
Here $\alpha$ is a mixing angle used to diagonalize the CP--even mass matrix
and $\beta$ is defined by $\tan\beta=v_2/v_1$ ($v_i$ is the VEV of the $i^{th}
$ doublet and $v_1^2+v_2^2=246^2$ GeV$^2$). In the MSSM, which is a constrained
version of the 2HDM (Model~ II), the angles $\alpha$ and $\beta$ are
related. For the models that we shall consider $\alpha$ and $\beta$ are 
independent. There exists a bound on $\tan\beta$ of 
\begin{equation}
\tan\beta\ge 1.25
\end{equation} 
obtained from considering the effects of the charged Higgs scalar on the 
$Z\to b\overline b$ decay \cite{Gross}.

For the MHDM there exist $N$ doublets with the VEVs ($v_i$)
obeying the relationship
\begin{equation}    
v^2=\sum_{i=1}^N v_i^2\;.
\end{equation}
The lightest CP--even mass eigenstate is a linear combination of the 
Lagrangian eigenstates, and can be written as
\begin{equation}
h_1= \sum_{i=1}^N X^*_{i1}\phi^0_i\;.
\end{equation}  
The parameters $X^*_{i1}$ originate from the mixing matrix for the CP-even
neutral Higgs sector defined by
\begin{equation}
\phi^0_i= \sum_{j=1}^N X_{ij}h_j\;.
\end{equation} 
They are therefore analogous to $\sin\alpha$ and 
$\cos\alpha$ in the 2HDM. 
Our convention is that $\phi^0_i$ is the real part of the neutral Higgs field
of the $i^{th} $doublet; $h_j$ are the mass eigenstates with $h_1$ 
taken to be
the lightest. The unitarity of the matrix forces the relationship
\begin{equation}
\sum_{i=1}^N |X_{i1}|^2=1\;.  
\end{equation}

The Yukawa couplings in the MHDM are not as correlated as in the 2HDM. 
Therefore a MHDM can always feign a 2HDM but might possibly
possess a distinctive signature. We shall now quantify the above statement.
In the 2HDM there exists the relation $v^2=v_1^2+v_2^2$. Since the strength of 
a Yukawa coupling is proportional 
to $v/v_i$, for small $v_i$ the coupling is significantly enhanced 
relative to
the minimal SM Higgs boson ($\phi^0$).  We see that if 
$v_1$ (say) is small, then $v_2\approx v$ is automatic. Therefore any Yukawa
coupling which is proportional to $v_2$ cannot be enhanced significantly 
compared to $\phi^0$. 
However in the MHDM one has the relation in Eq. (2). This is a weaker 
constraint on the $v_i$ and therefore it is entirely possible
that several $v_i$ are simultaneously small and hence a significantly
different phenomenology may be possible. The Yukawa couplings for a MHDM
are of the form 
\begin{equation}
vX^*_{i1}/v_i
\end{equation}
and so the enhancements possible in the 2HDM are attainable if $v_i\ll v$. 

We must next study the branching ratios (BRs) of the 2HDM and MHDM with the aim
of obtaining distinct signatures not possible for $\phi^0$ or $h^0$ of the MSSM.
For Higgs masses in the range of LEP2 ($M_{\phi^0}\le 100$ GeV) the only important
decays are to $b\overline b$, $\tau^+\tau^-$, $c\overline c$ and $gg$. For 
$\phi^0$ with $M_{\phi^0}=50$ GeV the ratios of the respective BRs are 
approximately \cite{Int}\footnote{QCD corrections are quite sizeable and have
been 
included.}:
\begin{equation}
b\overline b:\tau^+\tau^-:c\overline c:gg=0.87:0.08:0.03:0.02\;.
\end{equation}
For $M_{\phi^0}=100$ GeV one has:
\begin{equation}
b\overline b:\tau^+\tau^-:c\overline c:gg=0.80:0.09:0.03:0.07\;.
\end{equation}
Here the BRs for $b\overline b$ and $c\overline c$ are evaluated by summing
 the Feynman
diagrams in Figure 1, with the replacements $q=b$ and $q=c$ respectively. The 
virtual QCD corrections have not been displayed in the figures although they are
included in Eqs. (7) and (8).
The first two diagrams give the dominant contributions \cite{Zer}.
The BR for $gg$ represents the summation of all diagrams which
give light quark jets i.e. all diagrams in Figure 1 with $q=u$, $d$, $s$, and 
both diagrams in Figure 2. We note that decays via a quark loop proceed 
predominantly by a top quark loop.   
\begin{figure}[hp]
\begin{center}
\mbox{\mbox{\epsfig{file=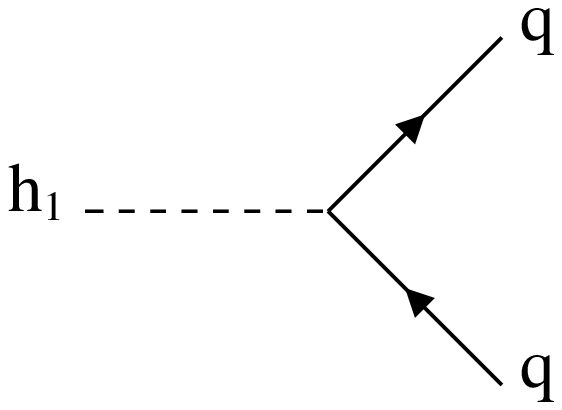,height=3cm}}
\mbox{\epsfig{file=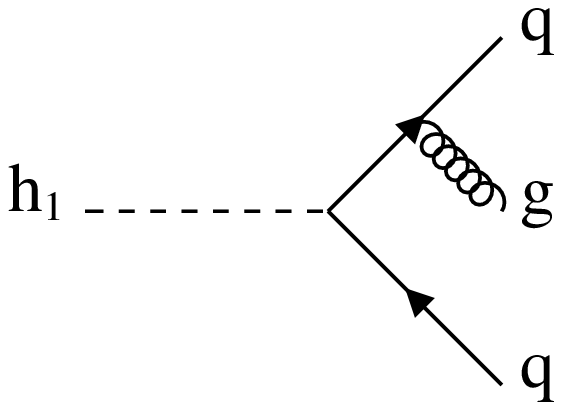,height=3cm}}
\mbox{\epsfig{file=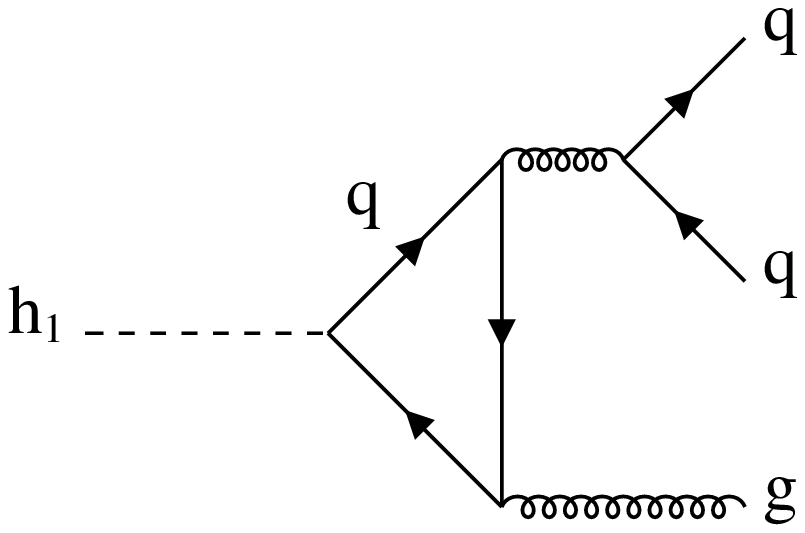,height=3.5cm}}}
\end{center}
\vspace{-5mm}
\caption{Decays of $h_1\to q\overline q$, $q\overline qg$.}
\label{Fig:Fig1}
\end{figure} 

\begin{figure}[hp]
\begin{center}
\mbox{\mbox{\epsfig{file=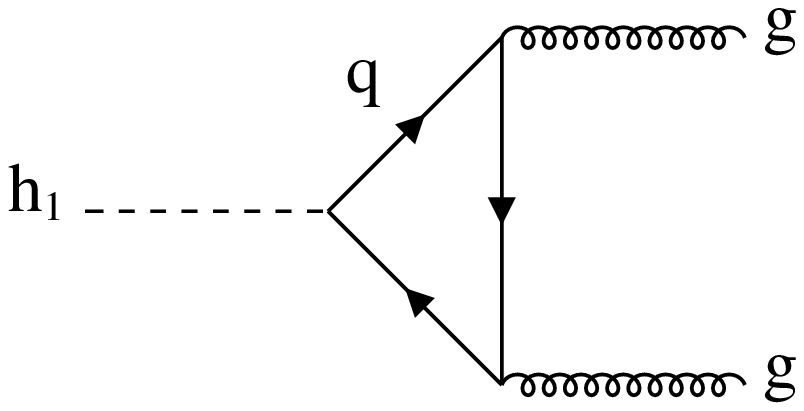,height=3cm}}
\mbox{\epsfig{file=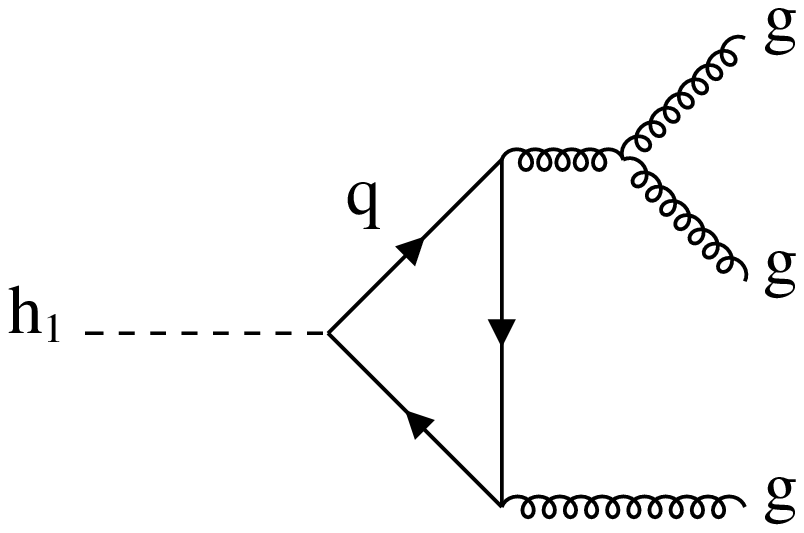,height=3.6cm}}}
\end{center}
\vspace{-5mm}
\caption{Decays of $h_1\to gg(g)$.}
\label{Fig:Fig2}
\end{figure} 
In Eq. (8) decays to $WW^*$ (where $*$ denotes an off--shell particle) are
at the $1\%$ level. It is apparent from Eqs. (7) and (8) that decays to
$b\overline b$ dominate, and this property allows light quark backgrounds to
the Higgs signal to be reduced by $b$--tagging. For $h^0$ one has
almost exactly the same BRs, although the $gg$ and $c\overline c$ decays are 
usually $\le 1\%$ unless $\tan\beta$ is large \cite{Int}; for large $\tan\beta$
they are both of the order $5\%$. Distinguishing $\phi^0$ from
$h^0$ provides a challenge for future colliders and is discussed
elsewhere \cite{Hab}. It is the aim of this paper to study the phenomenology
 of the 
2HDM and MHDM at LEP2, in the case of extreme BRs not possible for $\phi^0$ 
and $h^0$. From Table 2 one can vary the angles $\alpha$ and $\beta$ 
independently
to see if parameter spaces exist for such extreme BRs. We can summarize as 
follows: 
\begin{itemize}
\item[{(i)}] Model~ I$^{'}$: BR $(h_1\to \tau^+\tau^-$) large if $\tan\beta\gg
1$ and $\sin\alpha$ moderate\footnote{We take `moderate' to mean the angle
is approximately $\pi/4$.} to maximal. 
\item[{(ii)}] Model~ II$^{'}$: BR $(h_1\to \tau^+\tau^-$), BR $(h_1\to 
c\overline c$) and BR($h_1\to gg$) share domination for small $\sin\alpha$ 
and $\tan\beta$ moderate. 
\item[{(iii)}] Model~ II: BR($h_1\to c\overline c$) and  BR($h_1\to gg$)
share domination for small 
$\sin\alpha$ and $\tan\beta$ moderate.
\item[{(iv)}] Model~ I: Has exactly the same BRs as $\phi^0$.
\end{itemize}
One must be aware of the bound $\tan\beta\ge 1.25$ from Eq. (1) when varying 
the
angles. It is apparent that distinctive BRs {\sl are} possible in the
2HDM, the general requirement being that one of the angles has an extreme
value. Most noticeably the $b\overline b$ decays can be
suppressed and/or made negligible\footnote{We note that the width of the 
third diagram in
Figure 1 (with $q=b$) would be boosted if the coupling to up--type quarks is 
enhanced. However, this diagram has a very small width \cite{Zer} and so
 heavily suppressed
$h_1\to b\overline b$ decays are still possible.}. This is in contrast to 
$\phi^0$ and $h^0$
which will decay predominantly to $b\overline b$ for masses in range at LEP2 
-- see Eqs. (7) and (8). It is our aim to exploit these enhanced BRs at LEP2.
On first viewing it seems that BR($h_1\to \tau^+\tau^-) \gg 10\%$ 
would provide a clear signal of Model~ I$'$, since searching for $\phi^0\to 
\tau^+\tau^-$ in conjunction with $Z\to hadrons$ is one of the proposed 
techniques at LEP2 and gives a reasonable signal \cite{Int}. However, the 
production
cross--section $\sigma (e^+e^-\to Z^*\to Zh_1)$ is suppressed relative to that
of $\phi^0$ by a factor of $\sin^2(\beta-\alpha$) \cite{Gun}. Therefore we must 
investigate whether the parameter choices needed to obtain a distinctive BR 
simultaneously decrease the production cross--section at LEP2. We shall 
address this problem in the next section, as well as applying the proposed
search techniques for $\phi^0$ to the 2HDM and MHDM.           

\section{Phenomenology at LEP2}
The phenomenology of $\phi^0$ at LEP2 has been extensively covered in the 
literature, and recently in Ref. \cite{Int}. We shall be applying these 
various search techniques for the 2HDM and MHDM, and so will now give
a short review of them.

The dominant production process for $\phi^0$ if $M_{\phi^0}$ is in the range
of LEP2 ($M_{\phi^0}\le \sqrt s-M_Z$ GeV) is the Higgs--strahlung process
\cite{Pol} 
\begin{equation}
e^+e^-\to Z\phi^0
\end{equation}
in which $\phi^0$ is emitted from a virtual $Z$ boson. This cross--section
is well known and includes various loop corrections. We shall be interested
in the case of $\sqrt s=192$ GeV, since it is at this energy that the computer
simulations in  Ref. \cite{Int} have been carried out. The cross--section
for $\phi^0$ is $\sigma=0.95$ (0.40) pb if $M_{\phi^0}=60$ (90) GeV. For $M_{\phi^0}=100$ 
GeV, which is the mass generally
considered to be at the limit of the LEP2 range for $\sqrt s=192$ GeV,
$\sigma=0.1$ pb. Experimental simulations have been performed for the following
 event types:   

\begin{itemize}
\item[{(i)}] $Z\to e^+e^-;\;\mu^+\mu^-$, $\phi^0\to$ anything.
\item[{(ii)}] $Z\to \tau^+\tau^-$,  $\phi^0\to$ hadrons and vice versa.
\item[{(iii)}] $Z\to \nu\overline\nu$, $\phi^0\to$ hadrons.
\item[{(iv)}] $Z\to $ hadrons, $\phi^0\to b\overline b$. 
\end{itemize}

Events (iii) and (iv) require $b$--tagging to reduce the light quark
background, and since we shall be considering Higgs bosons with suppressed
$b\overline b$ decays these will not be relevant to our analysis.
Events of the form (i) are independent of the Higgs BRs, and require a 
pair of energetic leptons with an invariant mass compatible with $M_Z$.
The Higgs signal appears as a peak in the missing mass distribution.
Events of the form (ii) are characterized by two energetic, isolated $\tau$'s
associated with two hadronic jets. The existence of a Higgs boson would then
be observed as an accumulation around $(M_{\phi^0}$, $M_Z$) in the invariant 
mass
distributions of the $\tau$'s and jets. Table 3 (from 
Ref. \cite{Int}) shows the
simulated effective cross--section in the four channels for $M_{\phi^0}=90$ 
GeV and 
$\sqrt s=192$ GeV:
\begin{table}[htb]
\centering
\begin{tabular} {|c|c|c|c|c|} \hline
& $\phi^0l^+l^-$ & $\tau^+\tau^-q\overline q$ & $\phi^0\nu\overline \nu$ 
& $\phi^0q\overline q$  \\ \hline
ALEPH & 12:12 & 8:5 &  24:13  & 58:33     \\ \hline
DELPHI & 6:2 & 4:6 & 25:18 & 46:36      \\ \hline
L3 & 13:7  & -- & 28:21 & 36:23    \\ \hline
OPAL & 8:30 & -- & 17:13 & 34:36 \\ \hline
\end{tabular}
\caption{Effective cross--sections for signal (first number) and background
(second number) in fb, at LEP2 for $M_{\phi^0}=90$ GeV and $\protect\sqrt 
s=192$ GeV, 
taking into account acceptances and efficiencies (from Ref. $\protect
\cite{Int}$).}
\end{table} 

All the above analysis has been for $\phi^0$, with BRs given by Eqs. (7) and
(8). It is now our aim to extrapolate these results for use in the 2HDM
and MHDM. As mentioned in Section 2, these models (with the exception of Model
I) may possess vastly different BRs to those of $\phi^0$. 
However, the production cross--section for $h_1$ is suppressed relative to
that for $\phi^0$ (denoted by $\sigma_{\phi^0}$) by $\sin^2(\beta-\alpha$). 
A possible consequence of this suppression is that a very light $h_1$ 
$(M_{h_1}\approx 10$ GeV) would have
escaped detection at LEP1 if $\sin^2(\beta-\alpha)\ll 0.1$. This possibility
has recently been considered in Refs. \cite{Mar}, \cite {Kraw} and these
papers consider making use of non--suppressed production channels. For $h^0$
of the MSSM, $\sin^2(\beta-\alpha$) is close to one over essentially all
of the available parameter space and so there exists the lower bound 
$M_{h^0}\ge 44$ GeV \cite{Del}. Therefore the detection of a neutral 
CP--even Higgs boson
with a mass between 10 GeV and 40 GeV would indicate $h_1$. However, it
would be impossible to infer from which specific model it originates unless it 
displays a distinctive BR.  

To start our analysis let us now take the
case of Model~ I$'$ with a dominant BR $(h_1\to \tau^+\tau^-)$. One would 
therefore use detection channel (ii), and the event cross--section
(before applying efficiencies and acceptances) would be given by  
$\sigma_{\phi^0}\times \sin^2(\beta-\alpha)$ multipied by:
\begin{equation}
\left[{\rm BR} (Z\to q\overline 
q)\times {\rm BR} (h_1\to \tau^+\tau^-) + {\rm BR} (h_1 \to q\overline q)
\times {\rm BR} (Z\to \tau^+\tau^-)\right]\;. 
\end{equation}
Eq. (10) shows that the dominant term is the first since BR $(Z\to 
q\overline q)=70\%$, while BR $(Z\to \tau^+\tau^-)=3.4\%$. Now for $\phi^0$
we have the cross--section equal to 
\begin{equation}
0.086\times \sigma_{\phi^0}\;.
\end{equation}
For Model~ I$'$ this may be greatly enhanced, e.g. for BR $(h_1\to \tau^+\tau^-)
=80\%$ and BR~ $(h^0\to $ jets)=$20\%$ we have the following cross--section:
\begin{equation}
0.57\times \sigma_{\phi^0}\times \sin^2(\beta-\alpha)\;.
\end{equation}
Hence there could be
a considerable increase in the signal size for this channel, as much
as an order of magnitude. It is now
necessary to see if the values of $\beta$ and $\alpha$ needed to give a large 
BR $(h_1\to\tau^+\tau^-)$ cause a suppression in the $\sin^2
(\beta-\alpha)$ term. We note from the couplings in Table 2 that $\alpha\to 
\pi/2$ would further enhance $\tau^+\tau^-$ decays while suppressing 
the $b\overline b$ channel. However, from Eq. (12) we see that allowing
both $\alpha\to \pi/2$ and  $\beta\to \pi/2$ would cause strong suppression
from $\sin^2(\beta-\alpha)$ in the cross--section. We shall now derive
the value of $\alpha$ that maximizes the number of events.  

To obtain a large BR($h_1\to \tau^+\tau^-$) we showed in Section 2 that
$\tan\beta$ must be large and $\sin\alpha$ must be moderate to maximal. 
In Eq. (10)
the dominant term is the first one and so we shall neglect the second.
The first term is proportional to $\sin^2\alpha/\cos^2\beta$, and so 
the number of events in this channel depends on $f(\beta,\;\alpha$) defined
by:
\begin{equation}
f(\beta,\;\alpha)=\sin^2(\beta-\alpha)\times \sin^2\alpha/\cos^2\beta\;. 
\end{equation}
In Figure 3 we plot $f(\beta,\;\alpha$) with $\beta=1.5$ rads ($86^{\circ}$) 
since large BR($h_1\to \tau^+\tau^-$) requires small $\cos\beta$. 
\begin{figure}[th]
\begin{center}
{\mbox{\epsfig{file=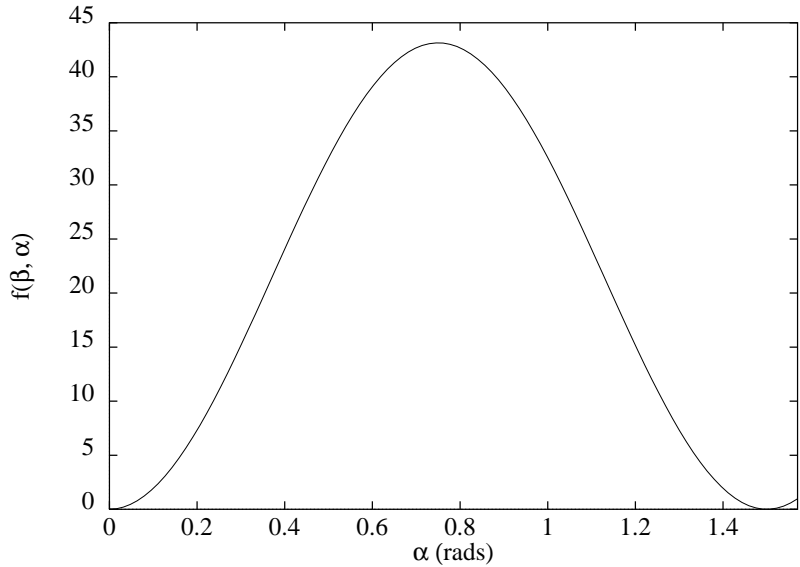,angle=-90,height=8.5cm}}}
\end{center}
\vspace{-5mm}
\caption{$f(\beta,\;\alpha)$ as a function of $\alpha$ for $\beta=1.5$ rads.}
\label{Fig:1}
\end{figure}

It is apparent from Figure 3 that moderate values of $\alpha$ ($\approx \pi/4$)
allow the largest values of $f(\beta,\;\alpha$). We see that 
$\alpha=\pi/4$ and $\beta\approx \pi/2$ would give $\sin^2(\beta-\alpha$)=0.5.
Therefore the suppression of the cross--section in Eq. (12) is not so severe, 
and the distinctive signature of a large number of 
$\tau^+\tau^-$ plus 2 jet events is certainly possible in Model~ I$'$.
Moreover, a large $\tau^+\tau^-$ peak 
will be centered at $M_h$ with almost no peak at $M_Z$, which is in contrast to 
the $\phi^0$ case which will have small $\tau^+\tau^-$ peaks centered at 
$M_{\phi}$
and $M_Z$ (since the two terms in Eq. (10) are of roughly equal magnitude). 
Hence, even if there is strong suppression from the $\sin^2(\beta-\alpha)$
term, the relative heights of the peaks will be different for $h_1$. 
However there may not be enough events to notice this pattern unless $M_{h_1}$ 
is considerably lighter than 90 GeV, i.e. when the cross--section is larger. 
Table 3
shows that for the ALEPH detector with 500 pb$^{-1}$ of integrated luminosity
one would expect the number of signal (background) events to be 4 (2). 
This is for $\phi^0$
with $M_{\phi}=90$ GeV. For $M_{h_1}=60$ GeV (say) the improved cross--section
would double the number of events.

For $h_1$ of Model~ I, Table 2 shows that its couplings to the fermions are
identical to those of $\phi^0$ and hence no distinct signature is possible. 
We recall that $h^0$ of the MSSM also feigns $\phi^0$ over a large parameter
space, but $h_1$ of Model~ I could have a suppressed production rate due
to the factor $\sin^2(\beta-\alpha$); hence a neutral CP--even Higgs boson
with identical BRs to $\phi^0$ but a suppressed cross--section and/or 
$M_{h_1}\le 44$ GeV would be indirect evidence\footnote{Of course, 
another 2HDM or a MHDM could mimic the BRs of $\phi^0$ for carefully
chosen couplings.} for Model~ I$'$. Our earlier work shows that
alternative distinctive signatures of Model~ I would  be a fermiophobic 
Higgs \cite{Ake1} and/or a light charged scalar ($M_{H^{\pm}}\le 80$ GeV) 
\cite{Ake}.  

For Model~ II$'$ we stated in Section 2 that the
decays $h_1\to \tau^+\tau^-$, $h_1\to gg$ and $h_1\to c\overline c$ would
share the domination for small $\sin\alpha$ and $\tan\beta$ moderate, i.e.
$h_1\to b\overline b$ may be heavily suppressed. The decay $h_1\to gg$ proceeds
via a top quark loop and so is boosted since the above parameter choices
enhance couplings to up--type quarks. From Eqs. (7) and (8) we see that
the ratio of the $\tau^+\tau^-$ partial width to the jet ($c\overline c$ and 
$gg$)
partial width would
be 8:5 for $M_{h_1}=50$ GeV, and 9:10 for $M_{h_1}=100$ GeV. Therefore in
contrast to Model~ I$'$ the $\tau^+\tau^-$ decays can never saturate the
total width (i.e. BR$\approx 100\%$). Hence it would appear that Model~ II$'$ 
could be mimicked by 
Model~ I$'$, although in this case Model~ I$'$ would have accompanying 
$b\overline b$ decays while Model~ II$'$ would be accompanied by 
$c\overline c$ and $gg$. If the $b\overline b$ decays are of sufficient
 magnitude then Model~ I$'$ would also register a signal in detection methods
 (iii) and (iv) (which require $b$--tagging), while Model~ II$'$ would not. 
Therefore it is possible that $h_1$ of Model~ II$'$ has a distinct
signature.

Model~ II is the structure that is most considered in the literature since
it is the form of the MSSM. As shown in Section 2 for suitable parameter
choices the $b\overline b$ channel can be suppressed and the decays 
$h_1\to gg$ and $h_1\to c\overline c$ would dominate. We do not believe that
this scenario has been considered before, and any detection technique that
requires $b$--tagging or $\tau^+\tau^-$ decays will fail. However, the 
missing mass technique is Higgs BR independent and so there is still a 
chance of a signal, provided that the cross--section suppression is not
too great. Figure 4 plots the suppression $\sin^2(\beta-\alpha$) as a function
of $\alpha$ for $\beta=\pi/4$ (moderate $\tan\beta$ is required for an
enhancement of $c\overline c$ and $gg$ decays)  
\begin{figure}[ht]
\begin{center}
{\mbox{\epsfig{file=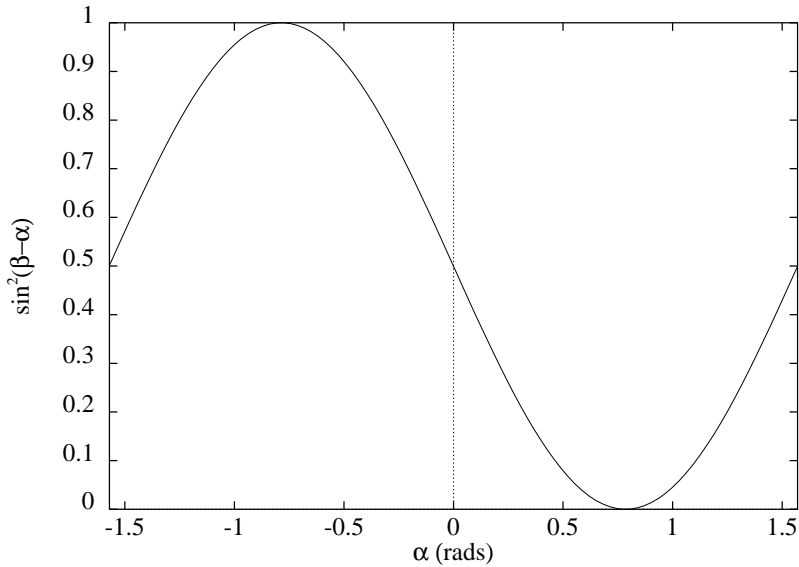,angle=-90,height=8.5cm}}}
\end{center}
\vspace{-5mm}
\caption{$\sin^2(\beta-\alpha)$ as a function of $\alpha$ for $\beta=\pi/4$
 rads.}
\label{Fig:2}
\end{figure}
and shows that for very small $\alpha$ (i.e. $\sin\alpha\approx 0$, which is
of interest to us)
the cross--section suppression is only of order 0.5. Therefore we anticipate
that a reasonable signal could be obtained for the scenario of $c\overline c$
and $gg$ decays dominating; see Table 3 for an indication of event numbers.
We stress that the non--observance of a Higgs signal in the other three
detection channels (which make use of $b\overline b$ and $\tau^+\tau^-$
decays), but a positive signal using the missing mass technique would be 
evidence for $h_1$ of Model~ II with $h_1\to c\overline c$, $gg$ decays 
dominating. Moreover, if such a signal is observed there would
then be motivation to search for charm rich jets originating from $h_1$. 
We conclude that Model~ II does possess a distinctive signature.

Having completed our account of the 2HDM at LEP2 we now address the MHDM.
As mentioned earlier, a MHDM can always feign a 2HDM but could it also possess
a distinctive signature? Concerning decay channels, the 2HDM can exhibit
extreme BRs for the decays to up--type quarks, down--type quarks and leptons.
Of course the MHDM can possess identical enhancements (Eq. (6)) but this is not 
distinctive. However, if the production cross--section of $h_1$ from the MHDM
has less suppression than $h_1$ from the 2HDM, as well as having an extreme
BR, then this would be a distinguishing trait. 
The coupling of the MHDM neutral scalars
to $Z$ (which is $\sin^2(\beta-~\alpha)$ in the 2HDM) is given by
\begin{equation}
{\sigma_{h_1}\over\sigma_{\phi^0}}= {|\sum v_iX_{i1}|^2\over v^2}\;.
\end{equation}
We recall that the scenario of an extreme BR in the 2HDM is accompanied by
a cross--section suppression of $\le 0.5$. From Eq. (14) we can show that
the suppression in the MHDM could be less. This is the case if a particular 
$v_i^2$
is much larger than the others ($v_i^2\to 246^2$ GeV$^2$), as well as 
possessing $|X_{i1}^2|$ close to 1. Therefore there exist parameter spaces for 
\begin{equation}
0.5\le {\sigma_{h_1}\over\sigma_{\phi^0}} \le 1. 
\end{equation}

\section{Conclusions}
We have studied the phenomenology of the lightest CP--even neutral 
Higgs boson ($h_1$) of a non--SUSY,
non--minimal Standard Model at LEP2. We considered the four versions of the 2HDM
and a general MHDM, and showed that $h_1$ could possess branching ratios (BRs)
significantly different to those of the minimal Standard Model Higgs boson 
$(\phi^0$) and $h^0$ of the Minimal Supersymmetric Standard Model;
 $\phi^0$ and $h^0$ decay 
predominantly to $b\overline b$ for masses in the range
of LEP2. It is assumed that $h_1$ is the only Higgs boson in range at LEP2.
The production cross--section of $e^+e^-\to Zh_1$ is suppressed relative
to that of $\phi^0$ by a factor $\sin^2(\beta-\alpha)$ in the 2HDM and an
analogous factor in the MHDM; a consequence of this suppression is that
a very light $h_1$ with 10 GeV $\le M_{h_1}\le$ 44 GeV may have escaped
detection at LEP1.
We showed that Model~ I$'$ may exhibit a large BR 
$(h_1\to \tau^+\tau^-$) which would register a clear signal. 
Model~ II may decay predominantly to $c\overline c$ and $gg$; this would provide
a distinctive signature since it could only be observed as missing mass
recoiling against a lepton pair (from $Z\to l^+l^-$), with no signal in the
other detection channels. For Model~ I, $h_1$ possesses the same BRs as $\phi^0$
and so does not have a distinctive signature. Model~ II$'$ may decay 
predominantly
to a mixture of $\tau^+\tau^-$, $c\overline c$ and $gg$; the enhanced
$\tau^+\tau^-$ channel could be mimicked by Model~ I$'$, but these latter decays 
would be accompanied by $b$ quark jets, in contrast to the 
light quark/charm rich jets which would accompany the enhanced $\tau^+\tau^-$ 
channel in Model~ II$'$. Therefore it is possible that Model~ II$'$
possesses a distinct signature. In the MHDM $h_1$ can 
feign any of the above extreme BRs 
but may be distinguished from $h_1$ of the 2HDM if its production 
cross--section is close to that of $\phi^0$.

\section*{Acknowledgements}
I wish to thank W.J. Stirling and C.J. Dove for reading the manuscript. 
This work has been supported by the UK PPARC.

\end{document}